\documentclass{emulateapj}
\def\kms{km\,s$^{-1}$\,}
\def\vs{$v_{S}$\,\,}
\def\etal{ et~al.\rm\,}
\def\sb{ergs cm$^{-2}$ s$^{-1}$ sr$^{-1}$\,}
\def\g292{G292.0+1.8\,}
\def\spitzer{{\it Spitzer\,}}
\def\chan{{\it Chandra}}

\tighten
\begin{document}
\submitted{Accepted by ApJ February 16, 2009}
\lastpagefootnotes

\title{\spitzer\, Spectroscopy of the Galactic
Supernova Remnant G292.0+1.8: Structure and Composition of the Oxygen-Rich Ejecta   }

\author{Parviz Ghavamian\altaffilmark{1}, John C. Raymond \altaffilmark{2}, William P. Blair\altaffilmark{3}, 
Knox S. Long\altaffilmark{1}, Achim Tappe\altaffilmark{2}, Sangwook Park\altaffilmark{4} and P. Frank Winkler\altaffilmark{5}
  }

\altaffiltext{1}{Space Telescope Science Institute, 3700 San Martin Drive, Baltimore, MD, 21218; parviz@stsci.edu; long@stsci.edu}
\altaffiltext{2}{Harvard-Smithsonian Center for Astrophysics, 60 Garden Street, Cambridge, MA 02138; jraymond@cfa.harvard.edu; atappe@cfa.harvard.edu}
\altaffiltext{3}{Department of Physics and Astronomy, Johns Hopkins University, 3400 N. Charles
Street, Baltimore, MD, 21218; parviz@pha.jhu.edu}
\altaffiltext{4}{Department of Astronomy and Astrophysics, Pennsylvania State University, 525 Davey Laboratory, University Park, PA, 16802; park@astro.psu.edu}
\altaffiltext{5}{Department of Physics, Middlebury College, McCardell Bicentennial Hall 526, Middlebury, VT, 05753; winkler@middlebury.edu}

\begin{abstract}

We present mid-infrared (5$-$40 $\micron$) spectra of shocked ejecta in the Galactic oxygen-rich 
supernova remnant \g292,  
acquired with the IRS spectrograph on board the \spitzer\, Space Telescope. The observations targeted two
positions within the brightest oxygen-rich feature in \g292.  Emission
lines of [Ne~II] $\lambda$12.8, [Ne~III] $\lambda\lambda$15.5,36.0, [Ne~V] $\lambda$24.3 and
[O~IV] $\lambda$25.9 $\micron$ are detected from the shocked ejecta.   In marked contrast
to what is observed in Cassiopeia A, no discernible mid-IR emission
from heavier species such as Mg, Si, S, Ar or Fe is detected in \g292.  We also detect a broad
emission bump between 15 and 28 $\micron$ in spectra of the radiatively shocked O-rich ejecta in \g292.
We suggest that this feature arises from either shock-heated Mg$_2$SiO$_4$ (forsterite) dust in the
radiatively shocked O-rich ejecta, or collisional excitation of PAHs in the blast wave of the SNR.  If the former
interpretation is correct, this would be the first mid-IR detection of ejecta dust in \g292.  A featureless dust continuum 
is also detected from non-radiative shocks in the circumstellar medium around \g292.  The mid-IR continuum from these structures, 
which lack mid-IR line emission, is seen in \chan\, images as bright X-ray filaments, is well described by a two-component silicate 
dust model.  The temperature of the hot dust component (M$_d\,\sim\,$2$\times$10$^{-3}$ M$_{\odot}$) is $\sim$115~K, while
that of the cold component (roughly constrained to be $\lesssim$\,3 M$_{\odot}$) is $\sim$35~K.  We attribute the hot component
to collisionally heated dust in the circumstellar shocks in \g292, and attribute the cold component
to dust heated by the hard FUV radiation from the circumstellar shocks.    
Using average O/Ne and O/Si mass ratios measured for a sample of ejecta
knots in the X-rays, our models yield line strengths consistent with 
mass ratios ${\rm M_{O}/M_{Ne}}$\,$\approx$\,3, ${\rm M_{O}/M_{Si}}$\,$\gtrsim$\,61 and ${\rm M_{O}/M_{S}}$\,$\approx$\,50. 
These ratios (especially the large O/Ne mass ratio) are difficult to reproduce with standard nucleosynthesis models of 
well-mixed supernova ejecta.  This reinforces the conclusions
of existing X-ray studies that the reverse shock in \g292\, is currently propagating into the
hydrostatic nucleosynthetic layers of the progenitor star, and has not yet penetrated the layers dominated by
explosive nucleosynthetic products. 

\end{abstract}


\section{INTRODUCTION}

The Galactic supernova remnant (SNR) \g292 is one of seven known SNRs whose optical spectrum is dominated by emission from metal-rich ejecta 
produced during a core collapse SN. The optical emission arises in dense knots of 
ejecta where the shocks have become radiative.  The prominent [O~I], [O~II] and [O~III]
emission lines observed in the spectra of these SNRs have earned them the title 'oxygen-rich supernova
remnants' (OSNRs).  Aside from \g292, the other known OSNRs include Cassiopeia A
(Minkowski 1957, Chevalier \& Kirshner 1978, 1979, Fesen \etal\, 2001) and Puppis A
(Winkler \& Kirshner 1985) in the Milky Way, 
N132D (Lasker 1978, 1980; Morse, Winkler \& Kirshner
1995; Morse \etal\, 1996; Blair \etal\, 2000) and 0540$-$69.3 (Kirshner \etal\, 1989; Serafimovich
\etal\, 2005) in the LMC, 1 E0102.2$-$7219 in the SMC (Dopita, Tuohy \& Mathewson 1981,
Blair \etal\, 2000) and the luminous, spatially unresolved OSNR in the irregular
galaxy NGC 4449 (Kirshner \& Blair 1980; Milisavljevic \& Fesen 2008).

More recently with the availability of high quality X-ray spectra of SNRs from \chan\, and {\it XMM}, core-collapse SNRs have been
identified purely from their line emission in X-ray spectra. The X-ray emission arises from 
faster shocks in lower density material, where the shocks remain in the non-radiative phase.  These objects include
purely non-radiative remnants such as B0049$-$73.6 in the SMC
(Hendrick, Reynolds \& Borkowski 2005) and E0103$-$72.6 in the SMC (Park \etal\,
2003).  The lack of optically-emitting ejecta knots in these objects indicates either that dense 
ejecta knots do not exist in these objects or that this material is not currently being encountered 
by the reverse shock.

Based on our current understanding, the core-collapse supernovae that produce OSNRs (i.e., those exhibiting oxygen line emission
in the optical) should leave behind rotating neutron stars (pulsars), but \g292\, is the only
Galactic remnant from this class found to harbor both an active pulsar
(Camilo \etal\, 2002, Hughes \etal\, 2003) and an associated pulsar wind
nebula (PWN) (Hughes \etal\, 2001).  Recent analyses of \chan\, data of \g292\,
(Park \etal\, 2002, 2004; Gonzalez \& Safi-Harb 2003) have shown that a prominent X-ray filament
stretching across the center of the OSNR seen in earlier {\it Einstein} observations (Tuohy, Burton \& Clark 1982)
is of normal composition. This suggests that \g292\, is interacting with
circumstellar material.  Due to the fact that is nearby, spatially resolved and exhibits all the expected features of a core-collapse
SN (O-rich optical and X-ray emission, active pulsar/PWN and apparent circumstellar interaction) it is critical to carry out as
many detailed studies of \g292\, in as many wavelength ranges as possible.
 
Recently the Infrared Spectrograph (IRS; Houck \etal\, 2004) on the \spitzer {\it Space Telescope (SST)} has provided a
major step forward in sensitivity and spectral resolution in the spectroscopy of objects in the mid-IR.  OSNRs, which exhibit
strong shock-excited line and continuum emission, are particularly well suited for study with the IRS.
The optical spectra of these remnants provide critical kinematic information from the Doppler shifts of strong
emission lines.  However, in the case of \g292\ and many of the other OSNRs, the optical spectra are dominated by few ions, 
making abundance measurements difficult.  In addition, most Galactic SNRs including \g292\, are located 
in the plane of the Galaxy where significant extinction 
prevents the detection of the UV lines expected from these objects.  These problems
are circumvented in the mid-IR, where emission lines from nearly all of the main burning products of core collapse SNe, including O, Ne, Mg, Si, S, Ar and Fe, 
are expected in the 5-40 $\micron$ bandpass of IRS.
In addition, the presence of multiple ionization stages from some species such as Ne, S, Ar and Fe
in this band and insensitivity of IR fine structure lines
to temperature help simplify the spectral analysis compared to optical and X-ray studies.  

The IRS also provides a sensitive tool for probing the properties of dust in both the ISM and supernova
ejecta.  Dust grains overrun by supernova shocks are progressively heated and destroyed downstream from 
the flow (Tielens \etal\, 1994), producing continuum emission that often peaks in the mid-infrared 
(Draine 1981, Dwek, Foster \& Vancura 1996).  Dust continuum is observed from fast ($\gtrsim$\,1000 \kms)
non-radiative shocks driven into the ISM and CSM of young SNRs (as in Kepler's SNR, see Blair \etal\, 2007),
and from radiative shocks (\vs\,$\sim$150-500 \kms) in dense ($n\,\sim\,$1000-10$^{4}$ cm$^{-3}$)
ISM/CSM and metal-rich ejecta, e.g., Cas A (Ennis
\etal\, 2005; Rho \etal\, 2008; Smith \etal\, 2008); E0102$-$72.3 (Sandstrom \etal\, 2008) and N132D (Tappe, Reach \& Rho 2006). 

With the above diagnostic capabilities of \spitzer\, in mind, we undertook a \spitzer\, project
to study the composition and physical state of the ejecta in \g292.  The goals of these observations
were to (1) measure the relative abundances of nucleosynthetic products within the shocked ejecta, and if possible, to
compare the results with theoretical models of massive stellar evolution to refine mass estimates
of the progenitor star; and (2) to investigate the emission from heated dust in \g292.  Throughout our discussion, IR wavelengths 
are shown in $\micron$ units, and optical wavelengths are given in \AA.
Below we describe the observational program undertaken and our analysis.

\section{OBSERVATIONS}

We observed \g292\, in a set of observations with the low resolution IRS module in 
early 2004 (Table~1).  The observations were performed in IRS Staring mode as part of the Cycle 
1 Guest Investigator program (PID 3598; P. Ghavamian, PI).
Details of the instrument configuration are listed in Table~1.  We observed two regions
in \g292, located in the upper and lower portions of the dominant oxygen-rich feature in \g292\,
known as the 'Spur' (Goss \etal\, 1979; Ghavamian, Hughes \& Williams 2005; Winkler \& Long 2006; hereafter WL06) (see Fig.~1).
The spectroscopic observations of the Spur included data from all four
modules, namely Short-Low (SL) 2nd order (5.2-8.7 $\micron$), SL 1st order (7.4-14.5 $\micron$),
Long-Low (LL) 2nd order (14.0-21.3 $\micron$) and LL 1st order (19.5-38.0 $\micron$).  The spectroscopic
resolution of the low resolution module lies in the range 64\,$\leq\, R ~\leq\,$128 in the wavelength
range 5.2\,$\leq\,\lambda\,\leq\,$38.0 $\micron$.  The spatial pixel scale of the SL modules is 1\farcs8 pixel$^{-1}$,
while that of the LL modules is 5\farcs1 pixel$^{-1}$.  

The positions of the slits for the observation dates are marked on the continuum-subtracted [O~III] image of \g292\, 
(WL06) in Fig.~1.  We list observational details such as the celestial coordinates
of each pointing, position angles of the slits and number of observing cycles in Table~1.  
The staring observation through each IRS module consisted of two dithered pointings wherein the
targeted feature is placed 1/3 of the distance along the length of the slit, then moved to a position
2/3 of the way along the length of the slit.  The resulting spectra overlap in a region approximately
38$\arcsec$ in length for the SL slits, and 112$\arcsec$ in length for the LL slits.  In the final
combined data, the object spectra are extracted from this overlap region, where the signal-to-noise
is highest.

We analyzed the \g292\, IRS spectra using the version S13.2.0 products of the
IRS calibration pipeline.  
Before extracting the spectra we cleaned the coadded the two-dimensional spectra from each nod into a single high signal-to-noise
spectrum.   For each set of pointings we first removed rogue pixels from the {\tt bcd}  images using the IRSCLEAN v1.7
algorithm\footnote{Developed by J. Ingalls and the Cornell IRS instrument team, and distributed by the \spitzer\, Science 
Center at {\sf http://ssc.spitzer.caltech.edu/archanaly/contributed/} }.  Next we shifted the second set of nods
by the length of each nod (11 pixels) to bring them into
alignment with the first set, then combined all of the aligned data together into a single high signal-to-noise
spectrum for each targeted position.  There are two sets of combined spectra for Spur positions 1 and 2
(effective integration time of 1200 s for SL1/SL2 and 2400 s for LL1/LL2).  We produced a combined uncertainty image for each
data set by taking the root sum square of the individual {\tt bcd} uncertainty images, then dividing the final 
uncertainty image by the square root of the total number of frames.  The combined bad pixel masks were
produced by adding the individual bad pixel masks for each {\tt bcd}.  After creating the final combined
images we extracted one-dimensional spectra using SPICE\footnote{The \spitzer\, IRS Custom Extraction Software is
distributed by the \spitzer\, Science Center at {\sf http://ssc.spitzer.caltech.edu/postbcd/spice.html} } v1.4.1.

Due to the extended nature of the targeted emission in \g292, the extraction and flux calibration of 
spectra required additional processing.  The spectral
extraction performed by the pipeline is optimized for point sources, with the extraction aperture expanding 
in the spatial dimension 
at longer wavelengths to account for widening of the instrumental PSF. The SNR emission in \g292,
however, is spatially extended.  If the emission is assumed to be uniform (admittedly a crude assumption
for \g292), then the emission gains from outside the extraction aperture compensate for losses at 
longer wavelengths, and the tapered extraction performed by the SPICE pipeline must be undone.  This is accomplished
by measuring the aperture loss correction function and the slit loss correction function for the
spectrum and, multiplying these functions by the object spectrum.  
The correction factors have been calculated 
as a function of wavelength for each IRS module by Tappe, Reach \& Rho (2006) from simulations of the \spitzer\, PSF.
These simulations indicate that the correction factors vary between 
0.6 and 0.8 over the full wavelength range covered by the two modules.  The final correction includes division of the
correction factors by the angular area covered by the extraction aperture.  Multiplication of this final correction
by a spectrum extracted in SPICE (units of Jy) gives a spectrum in units of Jy sr$^{-1}$.   

To obtain spectra of features in \g292\, we utilized the default extraction window used by SPICE for point source
extraction to obtain spectra in units of Jy.  We then multiplied these spectra by the extended source
correction function described above to obtain spectra in surface brightness 
units (here W m$^{-2}$ s$^{-1}$ $\micron^{-1}$).
After extraction in SPICE we loaded the resulting spectra into the SMART\footnote{SMART was developed by the IRS Team at
Cornell University and is available through the \spitzer\, Science Center at Caltech.} application (Higdon \etal\, 2004)
for analysis and post-processing.  

\section{ANALYSIS}

\subsection{DETECTED FEATURES}

We performed spectral extractions from the upper and lower portions of the Spur as shown in Figure~2,
corresponding to positions 1 and 2 listed in Table~1.  Of the data taken in the four low resolution modules,
all but the SL2 module (5.2-7.7 $\micron$) exhibit emission from either the SNR, its surrounding H~II
region, or both.   The combined, cleaned two-dimensional spectra are shown in Figures~3 and 4 without
background subtraction.  In these two-dimensional spectra, we detect atomic line emission from the following transitions: 
[Ne~II] $\lambda$12.8, [Ne~III] $\lambda$15.6, [S~III] $\lambda$18.7, [O~IV] $\lambda$25.9, [S~III] $\lambda$33.6 and [Si~II] $\lambda$34.8.
Less obvious in the two-dimensional spectra is [Ne~V] $\lambda$24.3, which is detected at the 15$\sigma$ and 23$\sigma$ levels
in the sky-subtracted, one-dimensional spectra of Spur1 and Spur2 (respectively).  A faint, diffuse emission feature is
detected at 11.3 $\micron$, consistent with a known polycyclic aromatic hydrocarbon (PAH) feature
at that wavelength (Allamandola, Tielens, \& Barker 1989).  A continuum, rising toward
longer wavelengths, is detected near the center of the LL1 and LL2 slits, consistent with dust emission
from localized clumps of material.  

Because our lines of sight pass well inside the projected edge of the SNR, the features detected in 
the IRS spectra of \g292\, arise from a combination of line/continuum
emission from photoionized gas, radiative shocks in ejecta, and non-radiative shocks in the lower 
density circumstellar medium/ISM around the SNR.  Although the gas phase
component of the first two
features can be traced at optical wavelengths via their forbidden line emission, the gas in the latter
feature is very hot (10$^{6}$-10$^{8}$~K) and ionized, and is only detectable at X-ray wavelengths.  On
the other hand, the heated dust component in photoionized regions, radiative shocks, and non-radiative shocks 
can produce continuum emission in the mid-infrared band (Tielens \& Hollenbach 1985; Vancura \etal\, 1994).  
Therefore, a careful comparison of the spatial information in the IRS spectra to the X-ray and optical images 
is critical for the identification and interpretation
of mid-infrared features detected in Figures~3 and 4.  As we show below, the dust continuum
observed from structures in the IRS slit can all be attributed to the belt-like structure of circumstellar
material observed
in the {\it Einstein} (Tuohy, Burton \& Clark 1982) and \chan\,
(Park \etal\, 2002, 2004, 2007) X-ray imagery of the OSNR.

There is significant structure evident along
the slit in the lines of [Ne~II] $\lambda$12.8, [Ne~III] $\lambda$15.6 and [O~IV] $\lambda$25.9,
and the emission from these lines rises sharply at the position of the ejecta in each
of the spectra.  These properties indicate strong emission from the Spur (ejecta) in O and Ne.
On the other hand, the [S~III] and [Si~II] emission is relatively uniform and appears to fill
the length of the slit, suggesting an association with the photoionized H~II region surrounding \g292\,
(this can be seen as the region of enhanced diffuse emission near the Spur in Figure~1).
The emission at $\lambda$25.9 may in principle be due to either
of [O~IV] or [Fe II].  These two spectral lines cannot be resolved from one
another at the spectral resolution of the LL module, but given the high ionization potential
of O$^{2+}$ (54.9) eV it is likely that the faint emission
seen near 25.9 $\micron$ is the [Fe II] $\lambda$25.99, while the bright emission
detected from the ejecta is [O~IV] $\lambda$25.9 $\micron$.  (As we show in the spectral
fits described in the next section, the centroid of the feature near $\lambda$25.9 $\micron$
is consistent with this line identification.)  

In addition to the above emission, there are also discrete, diffuse features detected
along the full length of the LL2 slit between 15 and 20 $\micron$ in all the unsubtracted spectra.  
The features blend together to form a plateau of emission.  Discrete PAH features are detected
at 16.4, 17.4 and 17.8 $\micron$ (consistent with photoionized emission from the overlying
H~II region) as well as a $H_2$ (0-0) S(1) emission feature at 17.03 $\micron$.  The
detection of $H_2$ in particular argues for a contribution to the mid-IR background from gas
along the line of sight unrelated to \g292.

\subsection{SKY VARIABILITY AND SUBTRACTION}

One of the main challenges in the interpretation of our \spitzer\, spectra of \g292\, is the
characterization and subtraction of the sky emission from our spectra of the O-rich shocks.  Since our observations
did not include a separate sky exposure, our best option was to measure the sky emission
from the off-object spectral order during each observation.  For example, in LL1 observations targeting
the Spur we extracted sky spectra from the LL2 channel, and vice versa for the LL2 observations.  We followed
a similar procedure for the SL spectra.  
We found that the Spur position 2 data provided the cleanest regions of sky emission, with the
least amount of contamination from dust continuum from circumstellar
material.  We extracted sky spectra from the Spur position 2 data, using the default point source extraction
aperture from SPICE as marked in Figure~2.  We then multiplied the spectra by the extended source correction 
factors described above to obtain the sky spectrum in surface brightness units.

The most obvious complication in our sky estimation 
is that each order samples a different section of sky emission (as shown in Figure~2).
In the optical, X-ray and now IRS data of \g292\, it is clear that the interior emission varies significantly
from location to location.
For example, it is evident from the [O~III] image in Figure~1 that the diffuse photoionized emission surrounding \g292\,
is distributed in a non-uniform manner: a swath of diffuse [O~III] emission extending 2\arcmin\, from the Spur
exhibits significantly higher surface brightness than the emission immediately to the north where the fast-moving
knots (FMKs) are observed (Ghavamian, Hughes \& Williams 2005; WL06).   More specifically, the
[O~III] surface brightness at the position where LL2 sky emission is sampled (Figure~2) is $\sim$70\% lower
than the diffuse [O~III] emission immediately surrounding the Spur, while the [O~III] sky emission sampled in the LL1 
spectrum is $\sim$20\% fainter than the diffuse emission around the Spur.  These differences likely correspond
to either variations in the density of the photoionized medium around \g292, or an enhancement in the flux
of UV radiation near the O-rich ejecta on the eastern side of the SNR, or some combination of both.  

Given the comparative faintness of the diffuse [O~III] emission in the sky slits relative to the emission
immediately surrounding the Spur, we can expect the mid-IR sky lines to also be intrinsically fainter
than those close to the Spur.  This results in under-subtraction of the line emission from the
mid-IR spectra of the Spur, an effect which is clearly seen in the extracted spectra from both
positions in the Spur (Figure~5).  Specifically, the faint [S~III]
$\lambda\lambda$18.7,33.6 and [Si~II] $\lambda$34.8 line emission in the spectra are likely
residuals from incomplete sky subtraction.

To extract the spectrum of Spur Position 1 we 
first obtained a spectrum from the center of the slit, where a strip of enhanced line and continuum
emission is seen in the two-dimensional LL1 and LL2 spectra (see Figure~3). 
We then subtracted the IRS sky spectrum obtained from the
SL1, LL2 and LL1 apertures marked in Figure~2 to obtain the final spectrum.
The line and continuum emission in the Position 1 
spectrum appear to arise from the same source (Figure~3), suggesting at first that we may have detected dust emission 
from radiative shocks in the Spur.  However, upon overlaying the IRS slit positions onto the [O~III] 
and \chan\, images of \g292\, it becomes clear that the situation is more complex. The extracted region coincides with both the circumstellar belt 
(seen in the \chan\, image in Figure~2) and the
radiative O-rich shocks in the Spur.  It is unclear which of these structures generates the observed continuum.  At mid-IR wavelengths 
the non-radiative shocks in the circumstellar belt are likely to produce strong continuum emission from shocked circumstellar/ISM
dust.  On the other hand, O-rich ejecta in the Spur are expected to produce strong line emission in the mid-IR.   However,
do they also contribute dust continuum emission?  With the given geometric projection of the Spur onto the circumstellar belt 
it is difficult to answer this question.

The answer to the above question may be found by inspection of the two-dimensional spectrum 
of Position 2 (Figure~4).  There, the radiatively shocked ejecta in the Spur (as detected in the LL1/LL2 observations)
can be isolated spatially from the circumstellar
X-ray belt, allowing us to better isolate emission from the two components.  From Figure~4 it is clear that at 
the top of the IRS slit the radiative O-rich shocks of the Spur (distinguished by strong [O~IV] $\lambda$25.9 
and [Ne~III] $\lambda$15.6 emission) are seen in projection on the circumstellar belt.   
Toward the bottom of the slit the two components
separate spatially from one another, revealing that the Spur is dominated in the mid-IR by line emission.  There is
no clear dust continuum detected from the Spur.  Extracting a spectrum from the continuum-free
portion of the Spur in Figure~4 and subtracting the same sky spectrum used for obtaining the spectrum of Position 1,
we obtained the spectrum for Position 2 shown in the lower panel of Figure~5.  
Save for the emission bump between 15 and 28 $\micron$, the sky-subtracted spectrum of Position 2 shows no
obvious underlying continuum.   The residual 15-28 $\micron$ bump 
indicates excess emission from this feature at the location of the Spur (we defer discussion of this
feature to Section 5).
In Table~2, we present measurements of the emission line surface brightnesses and velocities as measured from the
sky-subtracted data described above.  These line strengths will form the basis of our comparison to shock models
described in Section 4.

\subsection{CIRCUMSTELLAR DUST CONTINUUM}

Having separated the emission from the shocked radiative ejecta (line-dominated) and
the shocked non-radiative circumstellar belt (continuum-dominated), we now describe the properties
of each component.  We focus on the continuum from the belt observed in Position 1 (Figures~4 and 5).
After masking the emission lines in the Position 1 spectrum we fit the underlying continuum with a 
modified blackbody using the SMART application.   We fit two different dust types to the 
spectrum, one consisting of 0.1 $\micron$ silicate grains and one consisting of the $R_V$\,=\,3.1 grain mix 
suggested by Weingartner \& Draine (2001) for the Milky Way ISM. 
We obtained equally good matches from both dust models, although this is most likely due to the substantial flux
uncertainty in the data shortward of 22 \AA\, (i.e., SL1 and LL2 data), as well as relative uncertainties in the
flux calibration between channels.  The fit to the spectrum is primarily driven by the LL1 channel data, where
the flux errors are smallest.  
Therefore, aside from excluding a pure graphite composition for the dust grains, we cannot place strong
constraints on the relative concentrations of graphite and silicate grains in the circumstellar belt from 
our IRS spectrum of the belt.  

For the overall fit, we found it necessary to use two blackbody components to fit the spectrum, as shown 
in Figure 6: (1) a hot component ($T_d (hot)\,=\,$114$\pm$5~K for both assumed compositions)
to match the peak in the spectrum 
near 20 $\micron$, and (2) a cold component to match the excess IR continuum longward of 20 $\micron$
(Figure~6), with $T_d (cold)$\,=\,38$\pm$6~K for a pure 0.1 $\micron$ silicate distribution or
$T_d (cold)$\,=\,34$\pm$8~K for a Weingartner \& Draine (2001) distribution.   
While the error bars on the cold component from SMART are fairly small, the fact that only the tail of
this component lies in the observed range means that there is significant uncertainty in the characteristics
of this component.  However, clearly the presence of a cold component is indicated in our continuum fit
shown in Figure~6.  

From our dust model fits to the spectrum, we obtain a hot dust mass of (1.7$\pm$0.6)$\times$10$^{-3}$ $M_{\odot}$ 
and a cold dust mass of 3.1$\pm$1.4 $M_{\odot}$ for both a Weingartner \& Draine (2001) and 0.1 $\micron$ silicate dust composition.
The large dust mass obtained for the cold component is very sensitive to the temperature of this component,
which is itself very uncertain.  MIPS 70 $\micron$ imagery just obtained in our Cycle 4 \spitzer\, program will better
elucidate the characteristics of this component.

The temperature of the hot component and its close spatial correlation with the belt seen in the
\chan\, images strongly suggests that the continuum emission shortward of 20 $\micron$ arises from 
circumstellar dust overrun by the blast wave of \g292.  The need for two
grain temperatures indicates that there are either two different types of grains present within
the shocked circumstellar belt, or that the same species of grains exist at two distinct temperatures.  
Assuming the latter, along with the fact that the temperature of the cold component likely exceeds the 
temperature of 15-20~K expected from dust grains in an average interstellar radiation field (Li \& Draine 2001), 
then the two dust components can be attributed to two processes: collisional heating of grains behind the non-radiative shocks 
in the belt (hot component) and the heating of preshock
grains by FUV radiation from the radiative precursor in the belt (cold component).  The presence of faint
[O~III] $\lambda\lambda$4959,5007 from the circumstellar belt in \g292\, (Ghavamian, Hughes, \& Williams 2005) 
suggests that the shocks are becoming radiative there.  In that case, FUV radiation produced in the partially
formed cooling zone may be strong enough to heat the preshock dust in the belt up to $\sim$30~K.

\subsection{KINEMATIC PROPERTIES}

An interesting feature in our \spitzer\, spectra is the varying shape of the [Ne~II] $\lambda$12.8 line along the spatial length of the IRS slit.  Tracing
the shape of the [Ne~II] line from the top of the SL slit down to its bottom,
we can clearly distinguish an S-shaped variation in the shape of the line profile.  This pattern
can be seen at both observed positions in the Spur (Figures~3 and 4).
This pattern is not seen in the [Ne~II] sky emission, implying that
we are resolving velocity structure (bulk Doppler motions) in the shocked ejecta.  

We searched for a velocity shift in the centroid of the [Ne~II] emission between the two ends
of the slit in Spur Position.  We utilized a point source aperture extraction box to 
obtain two spectra, one from each end of the 38\arcsec\, overlap region between the two SL 1 nods (these
correspond to the top and bottom ends of the two-dimensional spectrum seen in Figure~3).  After
subtracting the sky emission from the two spectra,
we fit the [Ne~II] $\lambda$12.8 line profiles.  Although our fits
to the line did not reveal any additional broadening beyond
the instrumental resolution (3000 \kms\, at 12.8 $\micron$, or R\,$\approx$\,100), we did
detect a variation in the velocity centroid of the [Ne~II] $\lambda$12.8 line:
at the top (eastern) end of the slit the [Ne~II] $\lambda$12.8 velocity is blueshifted to a 
heliocentric velocity of $-$73$\pm$24 \kms, while at the bottom (western) end of the slit the line
is redshifted to a velocity of +290$\pm$13 \kms.  This result indicates that a velocity
gradient of $\sim$365 \kms\, exists in the Ne-rich ejecta of the Spur, consistent with results from
[O~III] $\lambda$5007 Fabry-Perot studies of the O-rich ejecta (Ghavamian, Hughes, \& Williams 2005).

\section{SHOCK MODELS}

The presence of high ionization lines such as [O~IV] $\lambda$25.9 and [Ne~V] $\lambda$24.3 in the
IRS spectra (Figure~5) suggests that the mid-IR emission from the Spur arises either partially or entirely
from radiative shock excitation.  To model the spectra we utilized the Raymond \& Cox (1985) numerical shock
code to predict line emission from radiative shocks in the metal-rich ejecta of \g292.  The version of the code that we used has been modified to handle the enormous cooling rates found in metal-rich
plasmas.  As noted in earlier studies of metal-rich shocks (Itoh 1981a, 1981b, 1986; Dopita, Binette
\& Tuohy 1984; Sutherland \& Dopita 1995; Blair et al. 2000) the radiative cooling time behind a shock driven into 
supernova ejecta is much shorter than the recombination time behind a shock propagating into a normal abundance medium.  As a result, the electrons
radiate away their energy rapidly within a thin ($\sim$10$^{13}$-10$^{14}$ cm) layer behind the
shock, reducing the temperatures to $\sim$100~K before the plasma can fully recombine.   Overall these properties mean that the
lack of ionization equilibrium found even in radiative shocks in cosmic abundance gas is much more pronounced
in metal-rich plasmas.

To model the line emission from \g292\, we created grids in shock parameter space, utilizing the same
version of the Raymond-Cox shock code used by Blair \etal\, (2000) to model the optical and UV
emission from the O-rich ejecta in N132D and E0102$-$72.3.  That version of the code calculates
the time steps behind the shock more carefully than the cosmic abundance version, providing more accurate predictions of the temperature, ionization state, and compression in metal-rich
shocks.   The modified code also includes a calculation of the charge exchange reaction
O$^{++}$\,+\,O$^{0}$\,$\rightarrow$\,2\,O$^{+}$, necessary for accurate estimation of the
oxygen ionization fractions.

\subsection{INPUT PARAMETERS AND MODEL SETUP}

We have combined both the fluxes of mid-IR emission lines (Table~2) and optical line fluxes measured in 
the Spur by WL06 to provide the best constraints on the physical conditions in the shocked ejecta.  The optical slit
position of WL06 runs E-W along the top of the Spur, nearly perpendicular to the IRS slit 
location in Spur Position 1 (Figure~1) of our \spitzer\, observations.  From East to West, the optical slit runs
first through low density S-rich ejecta (labeled Filament 1E by WL06) and then higher density O-rich
ejecta (labeled Filament 1W by WL06).  Filament 1W exhibits optical emission lines of O, Ne and S and
matches the location of Spur Position 1.
Thus, we focus our modeling efforts on the optical and 
mid-IR spectra of Spur Position 1/Filament 1W.

We constructed a grid of shock models assuming conservation of ram pressure ($n\,v^{2}_{S}$\,=\,const.) between the 
radiatively shocked ejecta and the hot non-radiative plasma behind the reverse shock.  To gauge the ram pressure
we examined the density-sensitive optical ratio of [S~II] 6716/6731 in the radiatively shocked ejecta.
For Filament 1W, WL06 found a [S~II] 6716/6731 ratio 
of 1.16, indicating that the bulk of the [S~II] emission arises from the postshock region where
$n_e\,\sim\,$300 cm$^{-3}$.  On the other hand, they found that the
temperature-sensitive ratio [O~III] (4959\,+\,5007)/4363 is approximately 17 (indicating an average
electron temperature $\sim$42,000~K and confirming that the [O~III] emission is primarily shock excited).  
These two results limit the range
of possible preshock densities and radiative shock speeds in the Spur, suggesting that preshock densities $\sim$0.5-10 cm$^{-3}$
and shock speeds $\sim$20-200 \kms\, would be appropriate.

The main goals of comparing our model predictions to the \spitzer\, observations are to determine whether the 
X-ray derived abundances are allowed by the optical/IR data and to determine whether additional abundance constraints 
can be imposed from our IRS spectra.  Park \etal\, (2004) measured mass ratios for O, Ne, Mg
and Si in four different ejecta knots.  \ Save for one ejecta feature in the NW (labeled 'Region
5' in their paper), the ejecta abundances in the other three ejecta knots are roughly consistent with one another.
We have averaged the mass ratios for these four knots (ratios from Table~2 of Park \etal\, 2004) to obtain an estimate
of the relative abundances in the radiative shocks in the Spur.  
From these ratios we estimated input abundances (by number) as a starting point for our models, obtaining
O:Ne:Mg:Si = 16.0:15.52:14.74:13.97.
Park \etal\, (2002, 2004) noted that they did not detect clear X-ray emission from elements heavier than Si in 
the ejecta of \g292.  Except for the S abundance, which can be constrained from the optical spectra of WL06,
the abundances of heavier elements in the ejecta of \g292\, such as Ar, Ca and Fe, remain unknown.  
The absence of  Ar lines ([Ar~II] $\lambda$6.99, [Ar~III] $\lambda$8.99) and Fe lines (such as [Fe~II] 
$\lambda$5.33, $\lambda$25.99 and [Fe~III] $\lambda$17.9) in our IRS spectra indicates a very low concentration of
these elements in the Spur.  The optical spectra also show no emission from these metals.
Similarly the abundance of carbon, a significant constituent of the ejecta, is unknown because the UV emission
lines from C are inaccessible due to the high extinction.  In the absence
of constraints on the abundances of C, Ar, Ca and Fe, we excluded these elements from the shock models.

The dependence of the optical and mid-IR fluxes on shock speed 
and preshock density must be constrained as well.   We normalized the ram pressure constraint $n\,v_{S}^{2}$\,=5000 
cm$^{-3}$ km$^{2}$ s$^{-2}$ by taking
$n$\,=\,0.5 cm$^{-3}$ at \vs =\,100 \kms.  The corresponding ram pressure is approximately 1.5$\times$10$^{-9}$
dyn cm$^{-2}$, comparable to the pressure of 5.9$\times$10$^{-9}$ dyn cm$^{-2}$ estimated
by Hughes \etal\, (2002) for an X-ray ejecta knot located just south of PSR J1124$-$5916 in \g292.   
With our ram pressure condition we ran a grid of shock models with intervals
\vs = (20,30,40,50,60,80,100,120,130,160,180,200) \kms\, and corresponding densities
$n$\,=\,(12.5, 5.5,3.1,2.0,1.4,0.8,0.50.35,0.25,0.2,0.15,0.13) cm$^{-3}$.

A number of other parameters must be set in the models.  We assumed photoionization equilibrium in the preshock
gas, i.e., we self-consistently calculated the
preshock ionization state of the ejecta by using the radiation output of the postshock gas.  
The degree of electron-ion temperature equilibration at the shock front is another free parameter.  
For our models, we set $T_e/T_i$\,=\,0.05.  The models follow the rise in temperature behind the
shock by Coulomb collisions and terminate the downstream calculation when the gas had cooled to 
300~K.\footnote{Currently the Raymond-Cox shock code does not include molecular cooling, which can, 
in principle, be important for $T\lesssim$1500~K.  However, the postshock gas in O-rich shocks
may cool rapidly enough to delay molecular formation until the gas reaches around 300~K.}  
Finally, following Blair \etal\, (2000) we set the value of
the magnetic parameter $B\,/\,n^{1/2}\,=\,$0.1 $\mu$G cm$^{3/2}$.

Inspection of the model grid is instructive.  For the adopted ejecta abundances and initial temperature
equilibration the slowest          
shocks ($\lesssim$40 \kms) produce strong [Ne~II] $\lambda$12.8 and [Si~II] $\lambda$34.8 in the mid-IR, and 
[S~II] $\lambda\lambda$6716, 6731 emission
in the optical.  In these cases the FUV radiation leaking upstream from the postshock flow
produces little preshock ionization.  Even after the gas crosses 
the shock the electron temperature is not high enough in these slow shocks to ionize the gas far beyond 
the second ionization stage.
This allows ions like Ne$^{+}$, Si$^{+}$ and S$^{+}$ to survive throughout the shock.  At higher shock speeds,
collisional ionization becomes more efficient.  The mid-IR line emission becomes less sensitive to
the preshock ionization state of the gas.  Emission from higher ionization stages dominates, with lines such as
[Ne~III] $\lambda\lambda$15.5, 36.0 and [S IV] $\lambda$10.5 reaching their peak around \vs\,=\,80 \kms\, and  
[O~IV] $\lambda$25.9 reaching a peak around 120 \kms.  The fastest shocks ($\gtrsim$150 \kms) are required to produce 
[Ne~V] $\lambda$24.3 emission.

\subsection{COMPARISON OF MODELING RESULTS TO OBSERVATIONS}

Judging by the IRS spectra of the Spur (Figures 3, 4 and 5; Table~2)
a wide range in shock speeds is apparently needed to simultaneously make the [Ne~II] $\lambda$12.8 lines
stronger than the [Ne~III] $\lambda$15.5 lines (implying \vs\,$\lesssim$40 \kms) and
produce detectable [Ne~V] $\lambda$24.3 and [O~IV] $\lambda$25.9 lines.
Following the treatment of Vancura \etal\, (1992) and Blair \etal\, (2000), we model the mid-IR
line emission by summing fluxes from the shock model grid between 20 \kms\, and 200 \kms\, with
a weighting term (\vs / 200)$^{-\alpha}$.  The power law index $\alpha$ is a free
parameter which can be adjusted to provide the best match with observed flux ratios.  Although $\alpha$ does not
have a rigorous physical definition, it may be interpreted as a measure of the preshock density distribution
within the ejecta.  A steep value of $\alpha$ emphasizes the slowest shocks and may reflect a small
range of density contrasts within a clump of ejecta. A shallower index, on the other hand, would
allow greater flux contributions from the intermediate and fast shocks, possibly reflecting
a wider range of density contrasts within the shocked ejecta.  Fitting the emission line ratios are
thus functions of both the relative abundances and power law index.  We estimated $\alpha$
by tracking relative fluxes from different ionization stages of the same element,
here the [Ne~II](12.8):[Ne~III](15.5):[Ne~V](24.3) and [O~III](4959 + 5007)/[O~III](4363) ratios,
before then varying relative abundances.

We combined the grid of models described above using various values of $\alpha$. We found
that $\alpha$\,=\,2.5 best matched the Spur Positions 1 and 2
[Ne~II](12.8):[Ne~III](15.5):[Ne~V](24.3) line ratios (Model A in Table~3).  
This index is considerably steeper than that obtained by Blair \etal\, (2000)
in modeling the radiatively shocked O-rich ejecta in N132D ($\alpha$\,=\,0.4) and 
E0102$-$72.3 ($\alpha$\,=\,0.5).  However, it is nearly identical to the index of 2.3 estimated
by Vancura \etal\, (1992) in their analysis of UV spectra from shocked interstellar clouds
in the SNR N49.  As discussed above, the steep index strongly weights the emission from the
slowest shocks, as necessary to produce [Ne~II](12.8)/[Ne~III](15.5) ratios
exceeding unity.  

The overall agreement between the $\alpha$=2.5 model grid and the observed optical and mid-IR flux ratios
for Spur Position 1 is reasonable (Table~3). Given the similarity between the mid-IR line ratios 
of Positions 1 and 2, the models essentially match the spectra from both positions.
The X-ray estimate of the O:Ne:Mg:Si ratios from Park \etal\, (2004) were able to match 
both the optical and mid-IR optical line ratios.  
Our models reproduce the optical line ratios of [S~II]
and [Ne~III] relative to [O~III] very well for Spur Position 1, and closely match the ratios of the brightest
lines in the \spitzer\, bandpass $-$ [Ne~II], [Ne~III] and [O~IV]. 
The modeled [Ne~II](12.8)/[Ne~III](15.5) ratio for Spur Position 2 is $\sim$20\% smaller than the observed
value.  Other important ratios
such as [Ne~III](15.5)/[O~IV](25.9) and [Ne~III](15.5)/[Ne~V](24.3) are more closely matched by the models,
though they are underpredicted as well ($\sim$15\%).
We found that by slightly steepening the index $\alpha$ from 2.5 to 2.7 (i.e., by increasing
the contribution from slower, denser shocks to the integrated mid-IR spectrum) we 
obtained [Ne~II](12.8)/[Ne~III](15.5)$\approx$2.5, matching the observed value of 2.7$\pm$0.24
in Position 2, while making [Ne~III](15.5)/[O~IV](25.9)$\approx\,$1.5, roughly consistent with
the observed value of 1.35$\pm$0.12.  There was minimal change in the other ratios.  The steeper
index for Position 2 is consistent with an ejecta distribution that is on average slightly denser
and clumpier.

As can be seen in Table~3, there are also significant discrepancies between the observed and predicted
optical ratios of [O~III]/[O~II] and [O~III]/[O~I] for Position 1.  In Model A the former ratio is
underpredicted by a factor of 2, while the latter is overpredicted by an order of magnitude.  
As noted by Itoh (1981a, 1986) and Blair \etal\, (2000) the difficulty in matching the [O~I] optical fluxes
is a well known shortcoming of existing models of radiative shocks in metal-rich ejecta.  The difficulty
is due to the fact that very little of the oxygen-rich gas behind the shock has recombined
to O~I before the postshock temperature drops below 1000~K.  This problem may
be partially solved by extending the range of shock speeds down to 10 \kms.  This modification
adds more [O~I] emission, thereby reducing the [O~III]/[O~I] ratio as needed.  However, it also significantly 
increases the [O~II] emission, worsening the disagreement between the observed and modeled [O~III]/[O~II] 
ratios.  Therefore, adding slower shocks does not entirely solve the line ratio problem.  Additional
steps such as increasing the assumed ratio $T_e/T_i$ at the shock front can further alleviate the problem
(see Model B in Table~3).  However, this creates disagreement with other line ratios such as
[Ne~II]/[Ne~III] and [Ne~III]/[O~IV] in the mid-IR.   Lastly, our models ignore the contribution
of photoionized unshocked ejecta clumps to the optical and mid-IR spectra of the O-rich knots.  Without
a detailed calculation (beyond the scope of this paper) the contribution of such clumps to the spectra
of O-rich knots in \g292\, is uncertain.  Taken together, all these difficulties illustrate the
sensitivity of the modeled spectra to the input shock parameters.  They remind us that such parameters as the mass
ratios of heavy elements, ejecta densities and range of shock speeds obtained from our models must be 
interpreted with caution.

As a consistency check we compared the surface brightness of [O~IV] $\lambda$25.9 predicted by Model A
(Table~3) with the observationally determined value at Position 1 of the Spur.  Summing the fluxes from
the plane parallel models and using the power law weighting in velocity described above, we obtained
a predicted [O~IV] $\lambda$25.9 flux of I$_{25.9}$({[O~IV])\,=\,
2.4$\times$10$^{-5}$ \sb, roughly twice the observed value (Table~2). 
Given our admittedly ad hoc use of a power law velocity distribution in calculating the total fluxes and
the model uncertainties described above, 
we consider this to be a reasonable agreement between the modeled and observed [O~IV] $\lambda$25.9 surface brightnesses.

We have also compared the surface brightness of [O~III] $\lambda$5007 predicted by the models
with the value observed in the optical by WL06.  Using the surface brightness of Filament 1W quoted by WL06
and applying an extinction correction using the relations of Fitzpatrick (1999), we obtained
a dereddened surface brightness $I_{5007}\,=\,$4.6$\times$10$^{-4}$ \sb.  Assuming that this number
is representative of the Spur emission sampled at Position 1 in our IRS spectra, we summed the emission
from our plane parallel shock models assuming the power law velocity distribution and obtained a predicted
surface brightness $I_{5007}\,=\,$1.6$\times$10$^{-3}$ \sb.  This result is also larger
than the observed value (by a factor $\sim$3).  Some of this difference may be ascribed to an overestimate of
the ejecta density by our model scaling, or it may be attributed to the uncertainty in extinction
correction, which ranges from a factor of 6.7 to 17.9 over the range in $E(B\,-\,V)$\,=\,0.6 to 0.9
(again assuming a Fitzpatrick (1999) extinction law).  However, we consider the overall agreement
between the observed and predicted [O~III] fluxes to be reasonable.

Although sulfur line emission has not been detected from the ejecta of \g292\, at X-ray wavelengths (Park \etal\,
2002, 2004), we were able to estimate or place limits on the abundance of this element from our
joint mid-IR/optical spectral analysis.  Starting from the averaged relative abundances
of O, Ne and Si obtained from Park \etal\, (2004), we progressively increased the
S abundance in the models until we matched the observed [S~II]/[O~III] flux ratio in the optical.  We obtained
a S abundance of 14.0 dex, with higher values producing unacceptably high [S~II]/[O~III] ratios and
resulting in enough collisional de-excitation of [S~II] $\lambda$6716 to reduce the [S~II] $\lambda$6716/$\lambda$6731
ratio below unity.  However, one remaining discrepancy is that our models predict a [S~III] $\lambda$18.7
flux that exceeds the upper limit on emission from that line in both IRS spectra of the Spur (Model A in Table~3).
The reason for the disagreement is unclear, but it could be ascribed to uncertainties in scaling 
between the optical and \spitzer apertures and the uncertain correction for strong photoionized [S~III] along
the IRS slit.

Turning to the abundances of magnesium and silicon, we find that the abundance of the former is 
unconstrained by our observations, since the [Mg~V] $\lambda$5.6 line is predicted by our models to lie nearly
an order of magnitude below the detection threshold at all shock speeds.  Therefore, we left the Mg abundance
at the average X-ray value of 14.74.  On the other hand, we were able to constrain the relative silicon abundance
in the Spur under the assumption that the [Si~II] $\lambda$34.8 emission in the IRS spectra is produced by radiative shocks
(rather than being the result of incomplete background subtraction).  We increased the abundance
of Si in our shock models (Model A) until we reached a Si abundance of 14.4 dex, above which the [Ne~III] $\lambda$15.5 / 
[Si~II] $\lambda$34.8 ratio dropped below the observed upper limit of 2.7 in Spur Position 1.  Therefore,
we estimate the upper limit on the Si abundance to be 14.4 in Position 1 of the Spur, corresponding
to $M_{O}/M_{Si}\,\gtrsim\,$23.

\section{DISCUSSION}

\subsection{DUST EMISSION}

The mid-IR spectrum of the radiatively shocked O-rich ejecta in \g292\, (e.g., the lower panel in Figure~5)
exhibits a faint, localized bump of emission between 15 $\micron$ and 28 $\micron$.  This contrasts strongly
with the situation in Cas A, where recent \spitzer\, IRS observations (Rho \etal\, 2008) have revealed continuum
emission from the ejecta extending across the IRS bandpass.  
The 19$-$23 $\micron$ continuum maps of Cas A obtained with IRS show a strong correlation
between the spatial distribution of emitting dust and that of ejecta-line emission from
[Ar~II] $\lambda$ 6.99, [Ne~II] $\lambda$12.8 and [O~IV] $\lambda$25.9.  This strongly suggests that
the dust observed in the mid-IR has condensed within the dense
ejecta in Cas A (Rho \etal\, 2008).   This clearly differs from the Spur in \g292, where,
despite the presence of strong Ne and O line emission, the only feature resembling a continuum 
is the broad emission bump between 15 and 28 $\micron$.   Summing the emission under the bump and excluding
emission lines, the surface brightness of the bump is $\approx$ 7.4$\times$10$^{-8}$
W m$^{-2}$ sr$^{-1}$.  If we adopt the this value as a 
measure of the dust continuum in \g292, then we estimate a line to continuum ratio $\sim$1 in the
mid-IR for the O-rich shocks in the Spur.
 
The efficiency of dust formation in supernova ejecta and the variables that
control it are currently not well known.  One 
source of uncertainty is the condensation efficiency of grains in the SN ejecta.  While
theoretical models based on classical nucleation theory predict that
0.1-0.3 M$_{\odot}$ of dust forms per SN (Kozasa, Hasegawa, \& Nomoto 1991; Todini \& Ferrara 2001),
infrared observations have turned up significantly lower dust mass estimates
in the ejecta of observed SNe.  For example, Sugerman \etal\, (2006) estimated that $\sim$0.02 $M_{\odot}$ of
dust had formed in the ejecta of SN 2003gd (although even this low number has been disputed; see
Meikle \etal\, 2007). The 0.02-0.05 M$_{\odot}$ of dust 
inferred to exist in the radiatively shocked ejecta of Cas A lies closer to (but still falls short of)
the theoretically predicted value.  If
the ejecta dust in Cas A and \g292\, are of comparable temperature and composition,
then the lack of a broad-band dust continuum in the mid-IR spectra of the latter might reflect a lower condensation 
efficiency in that OSNR.

The lack of strong dust continuum observed in the Spur of \g292\, may be similar to what was observed by
Tappe \etal\, (2006) in the O-rich ejecta of N132D in the Large Magellanic Cloud.  Their IRS spectrum,
which covered only the LL channel between 14 $\micron$ and 40 $\micron$, exhibits strong
[Ne~III] $\lambda$15.5 and [O~IV] $\lambda$ 25.9 lines, similar to the ejecta in \g292.  There is broad-band 
continuum emission present in the N132D spectrum.  However, given that Tappe \etal\, (2006) utilized background
regions lying outside the shell of N132D and that the ejecta are located near the center of this OSNR,
it is likely that the continuum emission detected in the ejecta spectrum of N132D is actually overlying 
dust continuum from shocked interstellar material along the line of sight.  If this is correct, 
then the line to continuum ratio of the ejecta knot in N132D may be similar that of the Spur in \g292.

A further parallel may be drawn between N132D and \g292\, using the HST FOS observations of N132D by Blair \etal\, (2000).
The optical and UV spectrum of the ejecta feature labeled N132D-P3 by Blair \etal\, (2000) was
found to have mass ratios of $M_{\rm O}/M_{\rm Ne}\,\approx\,$4.5 and $M_{\rm O}/M_{\rm Si}\,\approx\,$50.
These are very similar to the values we derived for the Spur.  Blair \etal\, (2000)
found that a low ram pressure, with $n\,=\,$1 cm$^{-3}$ at \vs\,=\,100 \kms\, was needed to reproduce the
UV and optical line ratios in the ejecta spectrum of N132D-P3, along with shock speeds in the range
of 30 \kms\, to 160 \kms.  These are again very similar to the conditions we found were needed to reproduce
the mid-IR and optical spectra of the Spur, though the velocity index $\alpha$ is only 0.4
in N132D, considerably less steep than the value of 2.5 found for the Spur.   Altogether it appears likely
that N132D and \g292\, are currently in similar stages of evolution, with the O-rich ejecta having
expanded to relatively low density and the reverse shocks having penetrated mostly the outermost
ejecta formed during the hydrostatic evolution of the progenitor.

\subsection{THE ORIGIN OF THE 15-28 $\micron$ BUMP} 

In both spectra of the Spur (Figure~5) there is a broad feature that peaks
near 17 $\micron$, then declines until it disappears near 28 $\micron$.  The feature resembles o the 15-20 $\micron$ bump
observed in the southeastern rim of N132D by Tappe \etal\, (2006; see Figure~6 in that paper).  Tappe \etal\, (2006) 
attributed the bump to in- and out-of-plane bending modes of the PAH C-C-C molecule, and suggested that the
PAH emission from the southeastern rim of N132D is generated by collisional
excitation behind the blast wave.  The smallest PAHs are preferentially destroyed behind the
shock, leaving the largest molecules de-hydrogenated in the form of large ($\gtrsim$10$^{4}$) clusters of carbon
atoms.  Since the smaller PAHs are responsible for the emission at 6.2, 7.7, 8.7 and 11.3 $\micron$,
the destruction of these molecules leaves the observed strong enhancement in the 15-20 $\micron$ / 11.3 $\micron$
flux ratio ($\sim$7) in N132D.

Recently Sandstrom \etal\, (2008) observed the OSNR E0102$-$72.3
with IRS and found a very similar emission bump in the spectrum of the radiatively shocked O-rich
ejecta of that remnant.  The bump in their spectrum extends out close to 30 $\micron$, much like
the feature detected in \g292.  The bump is prominent in the same spectra of  E0102$-$72.3 that show  the same emission lines from the radiatively shocked
ejecta of E0102$-$72.3 as we observe in \g292\, ([O~IV], [Ne~II], [Ne~III] and [Ne~V]). Instead of PAHs, Sandstrom \etal\, 
interpreted the bump as mid-IR emission arising from hot ($\sim$180~K) 
forsterite dust (Mg$_2$SiO$_4$) created within the SN ejecta.  Dust formation models (Nozawa \etal\, 2003) predict
that Al$_2$O$_3$ and Mg$_2$SiO$_4$ should form within 450 days after the SN explosion in region containing C-burning products of the O-Mg-Si layer in the outer
portions of the star.   If this is correct explanation for the dust  in the O-rich ejecta
of \g292\, should reflect the composition of the C-burning (hydrostatic) layers of
the progenitor star.  

Which of these suggestions is more likely to apply to the feature seen in \g292?  Observationally, the shape of the bump in \g292\ 
resembles that of  E0102$-$72.3 more than N132D.  The bump in the Spur 2 spectrum is considerably broader, extending
well beyond 20 $\micron$. In contrast, the feature
in the N132D spectrum is well confined between 15 and 20 $\micron$. Furthermore, the sky-subtracted Spur 2 spectrum
lacks the corresponding bending mode PAH features at 6.2, 7.7, 8.7 and 11.3 $\micron$ (the wavelength of the feature seen 
in the SL1 spectrum of Spur 2 is actually located at 10.5 $\micron$ (the wavelength of [S IV]) but is
a statistically insignificant ($\approx$1.5$\sigma$ detection).
If the Spur 2 bump is produced by dehydrogenated PAHs, then the 15-20 $\micron$ / 11.3 $\micron$ ratio
would have to be significantly larger ($\gtrsim$30) in \g292\, than in N132D.  Such a large ratio may be
be physically possible (if nearly all of the smaller PAH molecules have been destroyed behind
the blast wave), but a more detailed theoretical calculation is beyond the scope of this paper.  

From a more theoretical perspective, PAH emission is expected in situations where shocks encounter cold molecular gas since this is where they form.  
PAHs are a fairly plausible explanation in N132D, since the region surrounding N132D contains such a molecular cloud  (Tappe \etal\, 2006).   Clearly there is 
no such molecular material deep within \g292.  Therefore, if the 15-28 $\micron$
feature observed in the Spur spectra (Figure~5) arises from the excitation of PAH molecules, it would
have to originate in the circumstellar medium around \g292, which would probably only be feasible if the lines and dust emission arose from physically distinct regions.  
Thus an ejecta dust origin for the 15-28 $\micron$ bump in \g292\, is a much more appealing interpretation. However, it also must be reconciled with
the absence of [Si~II] $\lambda$34.8 in spectra of the Spur (Table~1).
A forsterite origin would require that most of the Si in the ejecta be depleted onto dust grains.  In the case of  E0102$-$72.3
(where there is similar lack of [Si~II]) this requirement would be less
stringent, due to the fact that E0102$-$72.3 is nearly 10 times farther away than \g292\,
and the IRS observations of the former encompasses a significantly larger fraction of the O-rich ejecta there than in the latter.  This
would increase the possibility that the mid-IR line emission and forsterite dust emission in E0102$-$72.3 arise from different locations,
and may relax the requirement of Si dust depletion in that OSNR.

\section{SUMMARY}

We have presented the first mid-infrared spectra of shocked O-rich ejecta in the 
Galactic SNR \g292.  These observations were obtained with the IRS spectrograph on \spitzer\, and
targeted the prominent structure known as `the Spur' on the eastern side of \g292.  The
only emission lines detected with certainty from the ejecta in the 5$-$40 $\micron$ range are 
those of oxygen ([O~IV] $\lambda$25.9)
and neon ([Ne~II] $\lambda$12.8, [Ne~III] $\lambda\lambda$15.5, 36.0 and [Ne~V] $\lambda$24.3).
While the IRS spectra of the Spur do not show the featureless continuum often associated
with dust emission, it does show a broad emission bump localized between 15 and 28 $\micron$.
This feature may be produced by either emission from dehydrogenated PAHs heated in
the forward shock of \g292, or it may be produced by Al$_2$O$_3$ or Mg$_2$SiO$_4$ (forsterite) dust grains in the
radiatively shocked O-rich ejecta.  It is similar to the bump seen recently in IRS spectra
of radiatively shocked ejecta in E0102$-$72.3 (Sandstrom \etal\, 2008).

Guided by a range of observed line ratios sensitive to
temperature, density, and relative abundances, we were able to constrain physical conditions in
the Spur using a grid of radiative shock models.  We calculated the spectra for radiative shocks 
with a range of speeds and preshock densities
subject to constant ram pressure constraints.  We then weighted the output fluxes with a power law in shock
speed to account for the range of density contrasts and shock speeds within the Spur.  We found
that shock speeds in the range 20$-$200 \kms, a power law index of $-$2.5 (preferentially weighting the slowest
shocks) and corresponding densities in the range 0.1$-$12.5
cm$^{-3}$ were needed to produce the range of low and high ionization emission lines observed
in the optical and mid-IR.   Furthermore, we found that simply by using average relative O:Ne:Si
abundances obtained from \chan\, spectroscopy of a handful of non-radiative ejecta knots by Park \etal\,
(2004) we were able to obtain good overall matches between our shock models and the observed
mid-IR and optical line ratios.  In the case of sulfur, which has not been clearly detected in the X-rays,
we were able to use the observed optical [S~II] $\lambda\lambda$6716, 6731 emission and the lack
of [S~IV] $\lambda$10.5 emission in the \spitzer\, data to estimate its abundance. 
Our results are O:Ne:Si:S = (16:15.52:13.97:14.2), with an
upper limit of 14.3 on the silicon abundance.  These correspond to ${\rm M_{O}/M_{Ne}}$\,$\approx$\,3,
${\rm M_{O}/M_{Si}}$\,$\approx$\,61 (limit $>$23) and ${\rm M_{O}/M_{S}}$\,$\approx$\,50.  While these
numbers are in broad agreement with those determined from X-ray observations, they were obtained
from shock models which ignore the contribution of photoionized, unshocked ejecta to the optical and
mid-IR emission in \g292.  More detailed and realistic models will be required to refine these 
estimates.

In addition to the radiative shocks in the O-rich ejecta, we have also detected mid-IR continuum
emission from the circumstellar belt observed in \chan\, X-ray images of \g292\ (Park \etal\, 2002; 2004).
X-ray analyses of the belt have indicated that it has abundances consistent with cosmic 
composition and point to excitation of circumstellar material in non-radiative shocks.    The mid-IR continuum
from the belt peaks at 20 $\micron$ and can be fit with a two-temperature dust model, with 
$T_d(hot)\,\approx\,$114~K and $T_d(cold)\,\approx\,$35~K.  The corresponding masses in the two components
are $M_d(hot)\,\approx\,$1.7$\times$10$^{-3}$ $M_{\odot}$ and $M_d(cold)\,\sim\,$3.1 $M_{\odot}$,
respectively, which we suggest are due to collisional heating of grains behind non-radiative shocks
in the belt (hot component) and the heating of preshock grains by FUV radiation from the radiative precursor
in the belt (cold component).  The continuum is equally well fit with
either a pure 0.1 $\micron$ silicate or a Milky Way (Weingartner \& Draine 2001) dust model.  

This work is based on observations made with the {\it Spitzer Space Telescope}, which is operated
by the Jet Propulsion Laboratory, California Institute of Technology under a contract with NASA.  Support
for the work of P. G. was supported by NASA through the \spitzer\, Guest Observer Program.

\clearpage

\begin{deluxetable*}{cccccccccc}
\tabletypesize{\tiny}
\tablecaption{Targeting Parameters for IRS Spectroscopy of the Spur in G292.0+1.8 (IRS Campaign 19)}
\label{tblobs}
\tablehead{
  \colhead{Position} & \colhead{AOR\tablenotemark{a} ID} & \colhead{$\alpha_{J2000}$} &
  \colhead{$\delta_{J2000}$} & \colhead{PA(\degr) (SL)}  & \colhead{SL Exp. Time (s)}  & \colhead{PA(\degr) (LL)}  &  \colhead{LL Exp. Time (s)}
& \colhead{Obs. Date} &
}
\startdata
Position 1  &  11254016
     &  $11^{\rm{h}}24^{\rm{m}}47\fs4$ & $-$59\arcdeg\ 15\arcmin\ 45\farcs1
     &  266.6   &  5$\times$240 &  183.0  &   5$\times$120  &  2005 Mar 24 \\
Position 2  &  11254272
     &  $11^{\rm{h}}24^{\rm{m}}44\fs1$ & $-$59\arcdeg\ 16\arcmin\ 10\farcs9
     &  266.6   &  5$\times$240  & 183.0  &  5$\times$120  &  2005 Mar 24  \\
\enddata
\tablenotetext{a}{Astronomical Observation Request}
\end{deluxetable*}

\begin{deluxetable}{cccccccc}
\tablecaption{Measured Surface Brightness of Emission lines in IRS Spectra of the Spur}
\label{tblobs}
\tablewidth{0pt}
\tablehead{
\colhead{ } & \colhead{Line ID}  &  \colhead{Wavelength ($\micron$)} & \colhead{Surf. Br. (10$^{-8}$ W m$^{-2}$ sr$^{-1}$)\tablenotemark{a}} &  \colhead{$V_{HELIO}$ (km s$^{-1}$)}\\
}
\startdata
{\bf Position 1}	&  &	&	&      &	&	&	\\
&  [S IV]  & 10.5  &  $<$ 0.25 &  \nodata   \\
&  [Ne II] &  12.8  &  3.55$\pm$0.07 &  +373   \\
&  [Ne III] & 15.6  &  1.7$\pm$0.09 &  +300   \\
&  [S III] & 18.7  &  $<$ 0.15\tablenotemark{b} &  $-$51   \\
&  [Ne V] & 24.4  &  0.12$\pm$0.01 &  +43   \\
&  [O IV] & 25.9  &  1.24$\pm$0.02 &  $-$51   \\
&  [S III] & 33.6  & $<$ 0.2\tablenotemark{b} &  +106   \\
&  [Si II] & 34.8  &  $<$ 0.63\tablenotemark{b} &  +117   \\
&  [Ne III] & 36.0  &  0.23$\pm$0.05 &  +385   \\
\\
{\bf Position 2} &   &       &       &      &        &       &       \\
&  [S IV] & 10.5  &  $<$ 0.68 &  \nodata   \\
&  [Ne II] & 12.8  &  4.21$\pm$0.14 &  +290   \\
&  [Ne III] & 15.6  &  1.55$\pm$0.13 &  +31   \\
&  [S III] & 18.7  &  $<$ 0.21\tablenotemark{b} &  $-$33   \\
&  [Ne V] & 24.4  &  0.11$\pm$0.01 &  $-$285   \\
&  [O IV] & 25.9  &  1.14$\pm$0.04 &  +11.5   \\
&  [S III] & 33.6  &  $<$ 0.2\tablenotemark{b} &  +168   \\
&  [Si II] & 34.8  &  $<$ 0.26\tablenotemark{b} &  +573   \\
&  [Ne III] & 36.0  &  $<$0.1  &  \nodata   \\

\enddata
\tablenotetext{a}{Equivalent to surface brightness in units of 10$^{-5}$ ergs cm$^{-2}$ s$^{-1}$ sr$^{-1}$ }
\tablenotetext{b}{These lines are most likely residuals from incomplete sky subtraction, so the fluxes
quoted here are upper limits on the flux from the radiatively shocked ejecta.  }
\end{deluxetable}

\begin{deluxetable}{lccccccc}
\tablecaption{Observed and Modeled Ratios of Emission Lines in the Spur }
\label{tblmodel}
\tablewidth{0pt}
\tablehead{
\colhead{Line Ratio}  &  \colhead{Position 1} & \colhead{Position 2}	&   \colhead{Model A\tablenotemark{a}} &   \colhead{Model B\tablenotemark{b}}  \\
}
\startdata
{\bf Mid-IR}	&	&	&	&	&      &  &	\\
$[$Ne II$]$(12.8)/$[$Ne~III$]$(15.5) & 2.1$\pm$0.1  & 2.7$\pm$0.24	&  2.2	&	0.4      \\
$[$Ne III$]$(15.5)/$[$Ne V$]$(24.3) & 14.1$\pm$1.4  & 14.1$\pm$1.74	&	 11.8  & 	11.7      \\
$[$S IV$]$(10.5)/$[$Ne II$]$(12.8) & $<$0.07  &		$<$0.16	  &	 0.04  &	0.23    \\
$[$Ne III$]$(15.5)/$[$O IV$]$(25.9) & 1.33$\pm$0.08  &	1.35$\pm$0.12	&	 1.3	&	0.7      \\
$[$Ne III$]$(15.5)/$[$Si II$]$(34.8) & $>$2.7  &  $>$6.0	&   11.1	&	4.1	  \\
$[$S III$]$(18.7)/$[$Ne III$]$(15.5) & $<$0.09  & 	$<$0.14 	&	0.22   &	0.05   \\
\\
{\bf Optical\tablenotemark{c} } &       &       &     &   	&	&	&       \\
$[$O III$]$(4959+5007)/$[$O II$]$(3727+3729) & 0.44 	&	\nodata  &  0.23	&	0.77   \\
$[$O III$]$(4959+5007)/$[$O II$]$(7325) & 16.7 	&	\nodata  &  4.55	&	19.5   \\
$[$O III$]$(4959+5007)/$[$O I$]$(6300+6363) & 13.1 	&	\nodata  &  149.2	&	4.9   \\
$[$O III$]$(4959+5007)/$[$O III$]$(4363) & 17.0  &	\nodata	&  18.7		&	20.0   \\
$[$S II$]$(6716+6731)/$[$O III$]$(4959+5007) & 0.13  &	\nodata  &	  0.13	&	0.16   \\
$[$Ne III$]$(3869)/$[$O III$]$(4959+5007) & 0.14  &  \nodata	&  0.14	&	0.13	   \\
$[$S II$]$(6716)/$[$S II$]$(6731) & 1.16  &  \nodata	&   1.2	&	1.3   \\
\enddata
\tablenotetext{a}{Model parameters are preshock density $n$\,=\,0.5 cm$^{-3}$ at 
shock velocity \vs\,=\,100 \kms, with $n\,v_{S}^{2}$\,=\,constant; $B/n^{1/2}\,=\,$0.1,
abundances in dex are O\,=\,16.00, Ne\,=\,15.52, Mg\,=\,14.74, Si\,=\,13.97
and S\,=\,14.0.  Fluxes from plane parallel shock models are combined with a power
law weighting factor as ($v_{S}$/200)$^{-\alpha}$, where 
we have assumed $\alpha\,=\,$2.5.   The range of shock velocities in the models
is 20\,$\leq$\,\vs\,$\leq$\,200 \kms, and $T_{e}/T_{i}$\,=\,0.05 is assumed at the shock front.  The Mg abundance is set to the averaged value
determined from X-ray observations and is not actually constrained by the optical/IR spectra.}
\tablenotetext{b}{Model B parameters are same as Model A, but with $\alpha$\,=\,0.4,  $T_{e}/T_{i}$\,=\,0.85,
and shock velocities in the range 10\,$\leq$\,\vs\,$\leq$\,130 \kms.}
\tablenotetext{c} {Optical line ratios for Spur 1 are those of Filament 1 W from 
the optical spectroscopy of WL06.  Ratios are quoted for $E(B\,-\,V)\,=\,$0.6 as estimated by WL06.
 }
\end{deluxetable}

\begin{figure}
\plotone{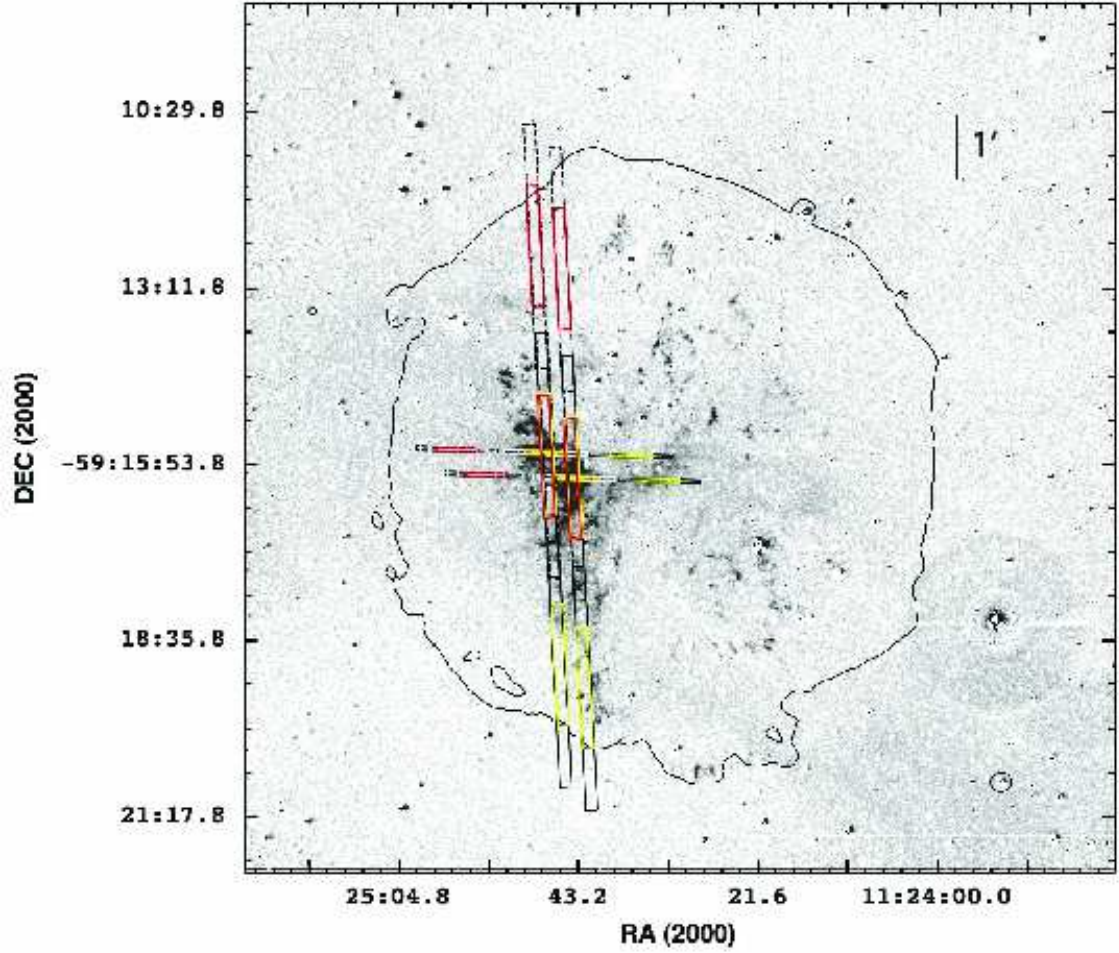}
\caption{
The continuum-subtracted [O~III] image of \g292\, (Winkler \& Long 2006) showing the
positions of the \spitzer\, IRS low resolution slits.  The outermost X-ray contour (0.4-0.95 keV emission
rom the 500 ks \chan\, LP observation of \g292; Park \etal\, 2007) is
shown marking the position of the blast wave.  East is to the left and North
is upward.  The O-rich Spur is covered by two IRS pointings, numbered positions 1 and 2 from
east to west (respectively).  Each slit is outlined according to the
following scheme: the first nod is marked as a solid black (or white) line, with the second nod as a dashed
black (or white) line. The overlap between the two nods where the signal-to-noise is maximized
in the combined data are marked in color: large red and yellow slits are LL1 and LL2, while
the small red and yellow slits are SL1 and SL2, respectively.  
}
\end{figure}

\begin{figure}[htp]
\plotone{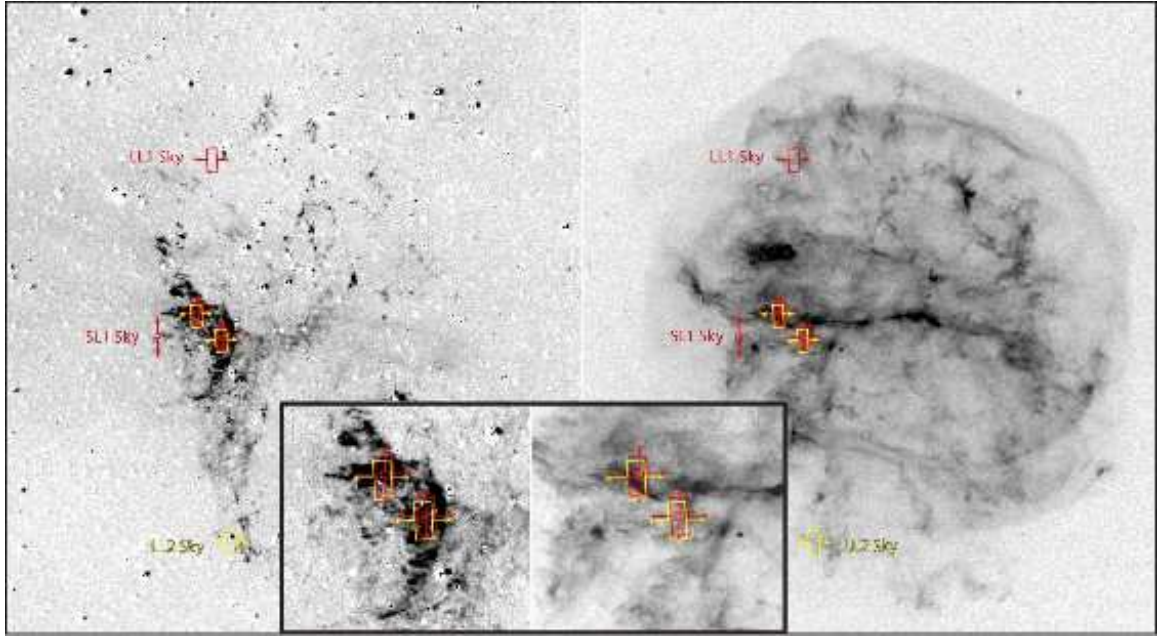}
\caption{ Same as Figure~1, but marked with the extraction boxes for the final
spectra (these exclude SL2, which did not yield obvious emission from \g292).   Left panel:
continuum-subtracted [O~III] image from Winkler \& Long (2006).   Right panel: the
\chan\, LP image of \g292\, (0.3-8.0 keV) (Park \etal\, 2007).  Field of
view of each image is approximately 8\farcm5 across.  The 
size of each box corresponds to the size of the spectral extraction
window at the central wavelength of each IRS module, assuming the point source extraction
aperture used by SPICE.  Colors indicate 1st order (red) and 2nd order (yellow).  
The arrows on each box indicate the wavelength dispersion
axis.  Dimensions are approximately 3.3 IRS pixels (= 6\arcsec)
along the spatial dimension and 3\farcs6 along the dispersion direction (= the slit
width) for SL1, and 4.5 pixels (= 23\arcsec) along the spatial dimension and
10\farcs5 along the dispersion direction (= the slit width) for LL 1.  
}
\end{figure}

\begin{figure}[htp]
\plotone{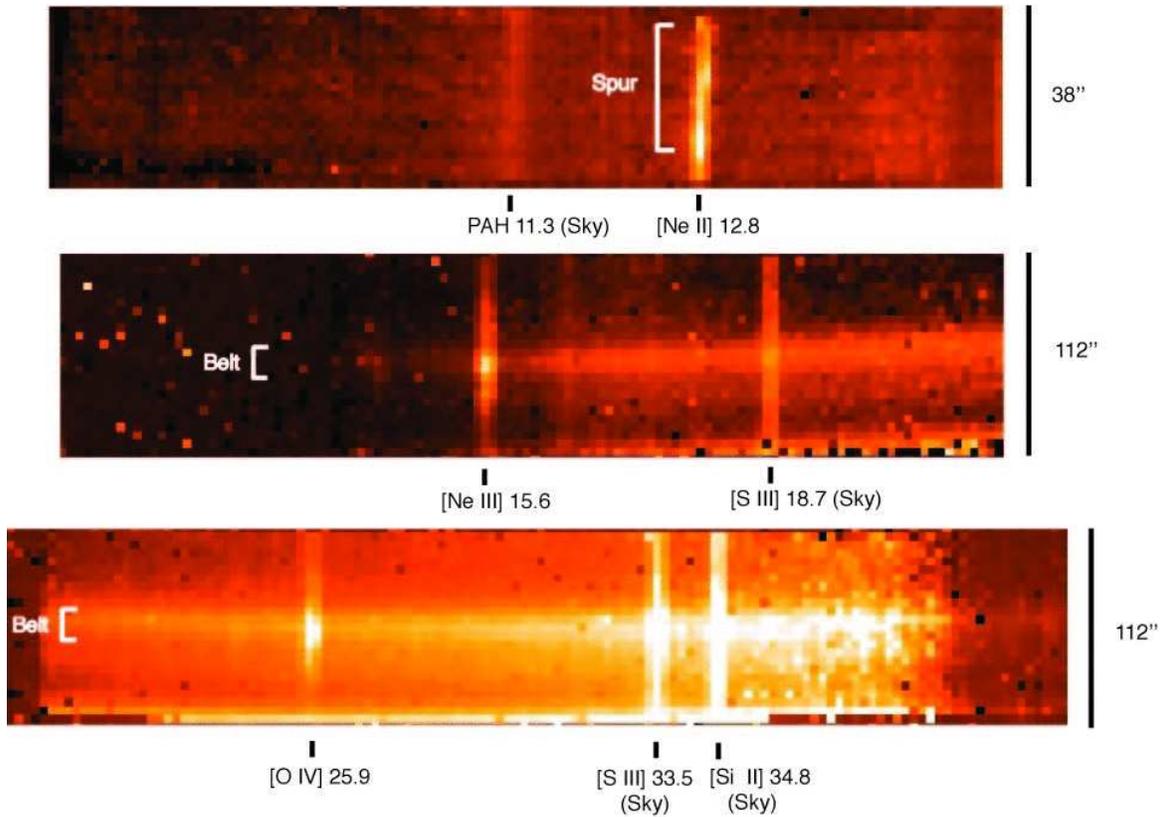}
\caption{The cleaned, combined two-dimensional spectra of Position 1 in the Spur, shown
from top to bottom for the SL1, LL2 and LL1 modules (the SL2 spectrum shows no discernible
emission and has been omitted).  No background subtraction has been
performed.  The positions of major features are marked.  East lies at the top of the SL 1
spectrum, and N at the top of the LL spectra (as seen in Figure~2).  Oxygen and neon line emission is
clearly detected from the Spur, while the continuum emission observed only near the center
of the slit arises from shock-heated dust in non-radiative shocks in the circumstellar belt. One-dimensional spectra were
extracted from the center of each slit shown in the figure, from regions in the SNR marked 
in Figure~2.  
}
\end{figure}

\begin{figure}
\plotone{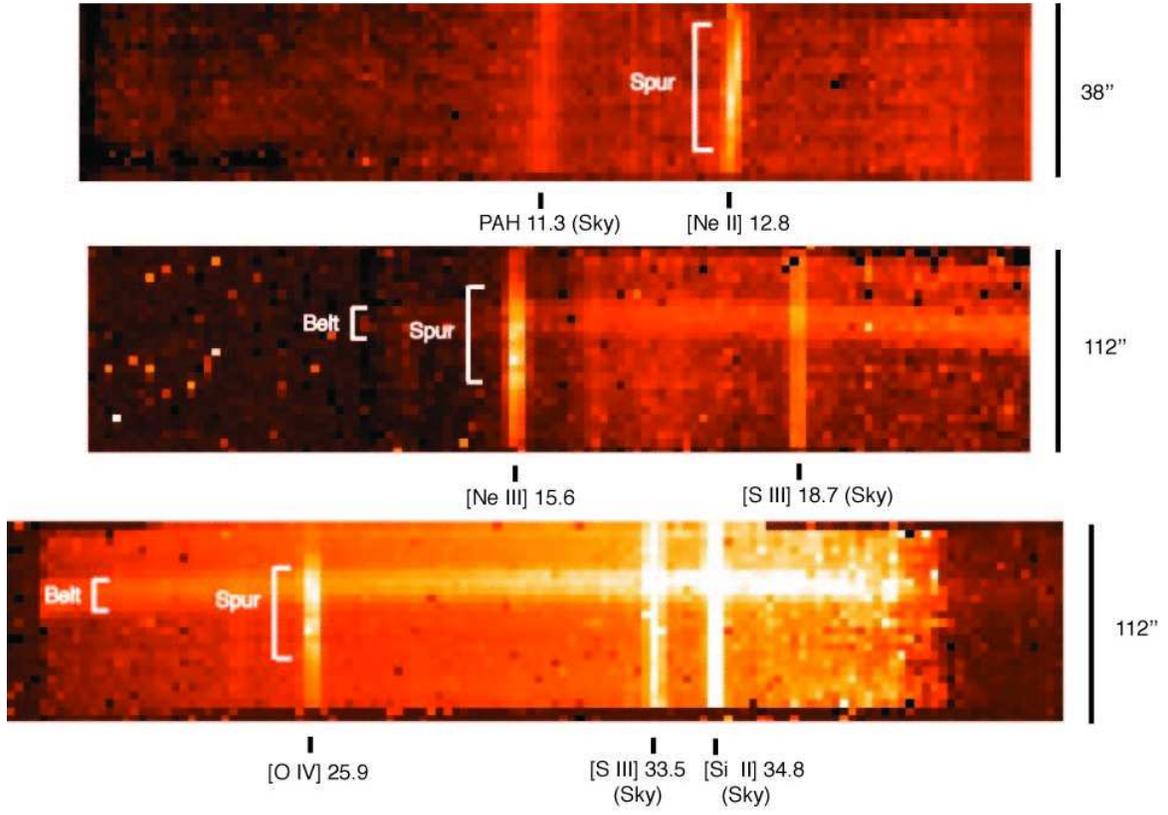}
\caption{Same as Figure~3, shown for Position 2 of the Spur.    No background subtraction has been
performed.  The positions of major features are marked.  East lies at the top of the SL 1 spectrum,
and N at the top of the LL spectra (as seen in Figure~2).  Dust continuum emission from the circumstellar belt
is again superimposed onto the emission line spectrum from radiatively shocked ejecta.  The Spur is
is detected in lines of O and Ne.  
}
\end{figure}

\begin{figure}
\plotone{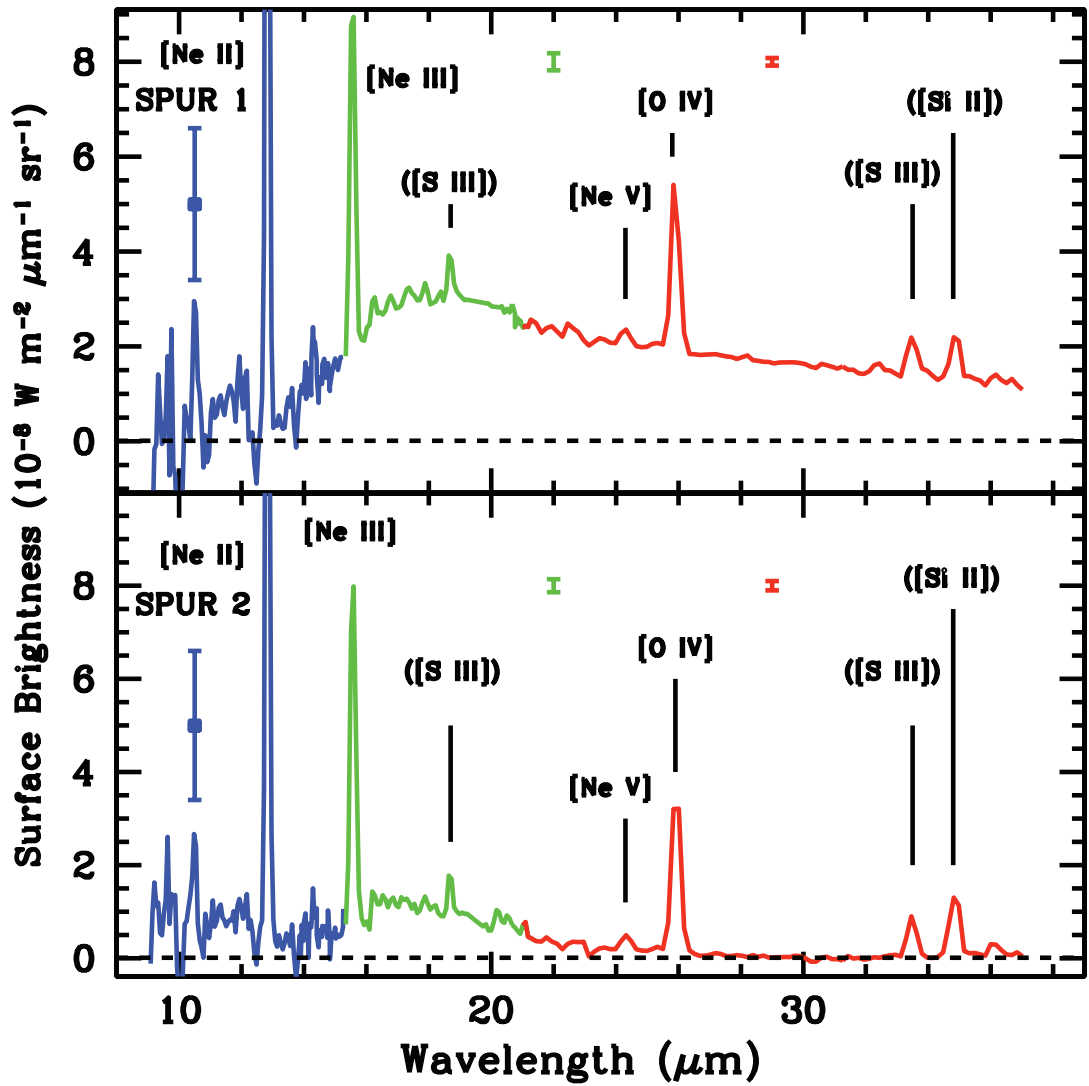}
\caption{
Background subtracted IRS spectra of two main features detected in \g292.  Identified emission
lines are marked.  The channel used to extract each portion of the spectrum is color coded
as blue (SL 1), green (LL 2) and red (LL 1).  The size of a typical flux error bar are marked
above the spectrum of each channel.  Labels marking the positions of features believed
to be residual sky emission are indicated in parentheses.    Top: the
spectrum of Spur Position 1 (emission lines + superimposed dust continuum from the belt).  Bottom: the spectrum
of position 2 in the Spur, where emission from ejecta has been isolated.  Note the broad emission feature
between 15 and 28 $\micron$, indicating either collisionally excited PAHs in the blast wave along the line
of sight or collisionally heated dust (Al$_2$O$_3$, Mg$_2$SiO$_4$) in the radiatively shocked
ejecta.
}
\end{figure}

\begin{figure}
\plotone{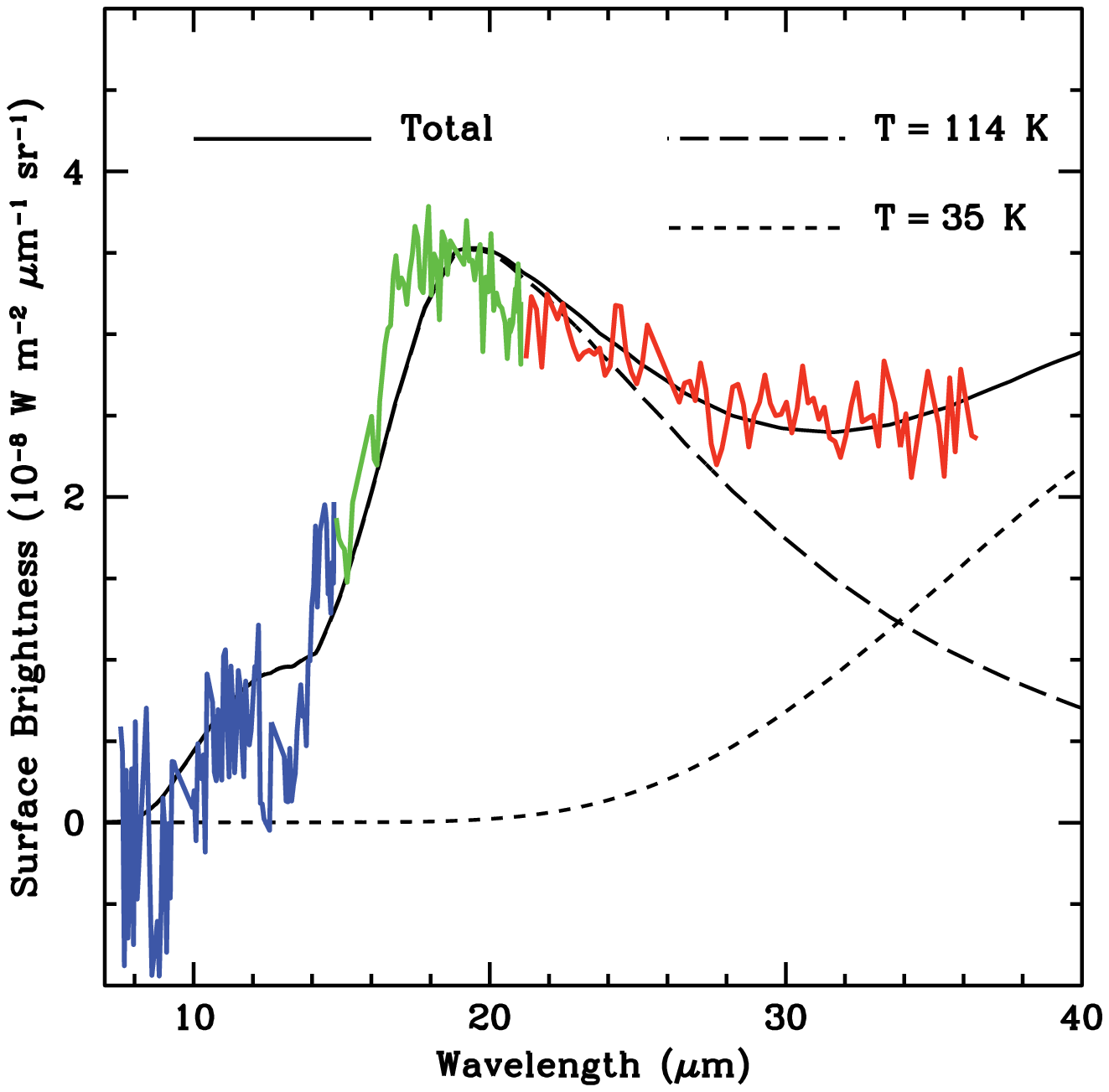}
\caption{Fit to the dust continuum from the circumstellar belt of \g292. 
The spectrum is that of Position 1 (top panel of Figure~5), but with emission lines from the overlying
shocked ejecta masked.   Data from   
the different channels are color coded blue (SL1), green (LL2) and red (LL1).  
The continuum is equally well matched with either a silicate dust model or Weingartner \& Draine (2001) model with
two components: a contribution from hot ($T_d\,\approx\,$114~K) dust at wavelengths
shortward of 30 $\micron$, and cold ($T_d\,\approx\,$35~K) dust dominant at wavelengths
beyond 30 $\micron$.  The fit shows an excess between 15 and 20 $\micron$, though the sizable
flux errors in the SL1 and LL2 data and systematic uncertainties in the 
IRS flux calibration between channels prevent a definitive interpretation
of the flux excess.
}\end{figure}


\begin{references}

\reference{A89}
Allamandola, L, J., Tielens, A. G. G. M., \& Barker, J. R. 1989, \apjs, 71, 733

\reference{B07}
Bianchi, S., \& Schneider, R. 2007, \mnras, 378, 973

\reference{Bl00}
Blair, W. P., \etal\, 2000, \apj, 537, 667

\reference{Bl07}
Blair, W. P., \etal\, 2007, \apj, 662, 998

\reference{B86}
Braun, R., Goss, W. M., Caswell, J. L., \& Roger, R. S. 1986, \aap, 162, 259

\reference{C02}
Camilo, F., \etal\, 2002, \apj, 567, L71

\reference{CK78}
Chevalier, R. A., \& Kirshner, R. P. 1978, \apj, 219, 931

\reference{CK79}
-----------, \apj, 233, 154

\reference{C05}
Chevalier, R. A. 2005, \apj, 619, 839

\reference{D81}
Dopita, M. A., Tuohy, I. R. \& Mathewson, D. S., \apj, 248, L105

\reference{D84}
Dopita, M. A., Binette, L., \& Tuohy, I. R. 1984, \apj, 282, 142

\reference{D87}
Dwek, E. 1987, \apj, 322, 812

\reference{D91}
Dwek, E., Foster, S. M. \& Vancura, O. 1996, \apj, 457, 244

\reference{E05}
Ennis, J., \etal\, 2005, \apj, 652, 376

\reference{F99}
Fitzpatrick, E. L. 1999, \pasp, 111, 63

\reference{G03}
Gaensler, B. M., \& Wallace, B. J. 2003, \apj, 594, 326

\reference{G05}
Ghavamian, P., Hughes, J. P., \& Williams, T. B. 2005, \apj, 635, 365

\reference{G79}
Goss, W. M., Shaver, P. A., Zealey, W. J. Murdin, P., \& Clark, D. H. 1979,
\mnras, 188, 357

\reference{H05}
Hendrick, S. P., Reynolds, S. P., \& Borkowski, K. J. 2005, \apj, 622 L117

\reference{Hi04}
Higdon, S. J. U., \etal\, 2004, \pasp, 116, 975

\reference{H04}
Houck, J. R., \etal\, 2004, \apjs, 154, 18

\reference{I81a}
Itoh, H. 1981, \pasj, 33, 1

\reference{I81b}
---- 1981, \pasj, 33, 521

\reference{I86}
---- 1986, \pasj, 38, 717

\reference{KB80}
Kirshner, R. P. \& Blair, W. P. 1980, \apj, 236, 135

\reference{K89}
Kirshner, R. P., Morse, J. A., Winkler, P. F. \& Blair, W. P. 1989, \apj, 342, 260

\reference{K91}
Kozasa, T., Hasegawa, H., \& Nomoto, K. 1991, \aa, 249, 474

\reference{L78}
Lasker, B. M. 1978, \apj, 223, 109

\reference{L80}
-----------, 1980, \apj, 237, 765

\reference{LD01}
Li, A., \& Draine, B. T. 2001, \apj, 554, 778

\reference{M07}
Meikle, W. P. S., \etal\, 2007, \apj, 665, 608

\reference{MF07}
Milisavljevic, D., \& Fesen, R. 2008, \apj, 677, 306

\reference{M57}
Minkowski, R., 1957, IAU Symp. No. 4, Cambridge Univ. Press, p. 107

\reference{M95}
Morse, J. A., Winkler, P. F. \& Winkler, \aj, 109, 2104

\reference{M96}
Morse, J. A., \etal\, \aj, 112, 509

\reference{M79}
Murdin, P., \& Clark, D. H. 1979, \mnras\ 189, 501

\reference{N03}
Nozawa, T., Kozasa, T., Umeda, H., Maeda, K. \& Nomoto, K. 2003, \apj, 598, 785

\reference{P02}
Park, S., \etal\, 2002, \apj, 564, L39

\reference{P03}
Park, S., \etal\, 2003, \apj, 598, L95

\reference{P04}
Park, S., \etal\, 2004, \apj\, 602, L33
 
\reference{P07}
Park, S., \etal\, 2007, \apj\, 670, L121
 
\reference{P04}
Peeters, E., Mattioda, A. L, Hudgins, D. M., \& Allamandola, L. J. 2004, \apj, 617, L65

\reference{R06}
Reach, W. T., \etal\, 2006, \aj, 131, 1479

\reference{R06b}
Reach, W. T., Rho, J., \& Jarrett, T. H., \apj, 618. 297

\reference{R07}
Rho, J. H., \etal\, 2008, \apj, 673, 271

\reference{S08}
Sandstrom, K., Bolatto, A. D., Stanimirovic, S., van Loon, J., \& Smith, J. D. 2008,
ApJ, submitted (astroph/0810.2803)

\reference{S05}
Serafimovich, N. I., Lundqvist, P., Shibanov, Yu. A., \& Sollerman , J. 2005, AdSpR, 35, 1106

\reference{S08}
Smith, J. D. T., \etal\, 2008, \apj, submitted (astroph/0810.3014)

\reference{S05}
Stanimirovic, S., \etal\, 2005, \apj, 532, L103

\reference{S06} 
Sugerman, B. S., \etal\, 2006, Science, 313, 196

\reference{S95}
Sutherland, R. S., \&  Dopita, M. A. 1995, \apj, 439, 365

\reference{T07}
Tappe, A., Rho, J., \& Reach, W. T. 2006, \apj, 653, 267

\reference{T85}
Tielens, A. G. G. M., \& Hollenbach, D. 1985, \apj, 291, 722 

\reference{T01}
Todini, P., \& Ferrara, A. 2001, \mnras, 325, 726

\reference{T82}
Tuohy, I. R., Burton, W. M. \& Clark, D. H. 1982, \apj, 260, L65

\reference{V94}
Vancura, O., Raymond, J. C., Dwek, E., Blair, W. P., Long, K. S. \& Foster, S. 1994, \apj, 431, 188

\reference{V00}
Van Kerckhoven, C., \etal\, 2000, \aap, 357, 1013

\reference{W01}
Weingartner, J. C., \& Draine, B. T. 2001, \apj, 548, 296

\reference{W06}
Williams, B. J., \etal\, 2006, \apj, 652, L33

\reference{WK85}
Winkler, P. F. \& Kirshner, R. P. 1985, \apj, 299, 981

\reference{WL06}
Winkler, P. F., \& Long, K. S. 2006, \aj, 132, 360 [WL06]



\end{references}
\end{document}